\documentclass[8.5pt,twoside,twocolumn]{article}
\usepackage[super,sort&compress,comma]{natbib} 
\usepackage{mhchem}
\usepackage{times,mathptm}
\usepackage{sectsty}
\usepackage{balance} 

\usepackage{graphicx} %eps figures can be used instead
\usepackage{lastpage}
\usepackage[format=plain,justification=raggedright,singlelinecheck=false,
font=small,labelfont=bf,labelsep=space]{caption} 
\usepackage{fancyhdr}
\DeclareTextSymbol{\degre}{OT1}{23} % for degrees
\newcommand{\latin}[1]{\textit{#1}} % for latin words, like ``via'', ``i.e.''...
\begin{document}

% \thispagestyle{plain}
% \fancypagestyle{plain}{
% \fancyhead[L]{\includegraphics[height=8pt]{headers/LH.pdf}}
% \fancyhead[C]{\hspace{-1cm}\includegraphics[height=20pt]{headers/CH.pdf}}
% \fancyhead[R]{\includegraphics[height=10pt]{headers/RH.pdf}\vspace{-0.2cm}}
% \renewcommand{\headrulewidth}{1pt}}
% \renewcommand{\thefootnote}{\fnsymbol{footnote}}
% \renewcommand\footnoterule{\vspace*{1pt}% 
% \hrule width 3.4in height 0.4pt \vspace*{5pt}} 
% \setcounter{secnumdepth}{5}

\makeatletter 
\def\subsubsection{\@startsection{subsubsection}{3}{10pt}{-1.25ex plus -1ex
minus -.1ex}{0ex plus 0ex}{\normalsize\bf}} 
\def\paragraph{\@startsection{paragraph}{4}{10pt}{-1.25ex plus -1ex minus
-.1ex}{0ex plus 0ex}{\normalsize\textit}} 
\renewcommand\@biblabel[1]{#1}            
\renewcommand\@makefntext[1]% 
{\noindent\makebox[0pt][r]{\@thefnmark\,}#1}
\makeatother 
\renewcommand{\figurename}{\small{Fig.}~}
\sectionfont{\large}
\subsectionfont{\normalsize} 

% \fancyfoot{}
% \fancyfoot[LO,RE]{\vspace{-7pt}\includegraphics[height=9pt]{headers/LF.pdf}}
% \fancyfoot[CO]{\vspace{-7.2pt}\hspace{12.2cm}\includegraphics{headers/RF.pdf}}
% \fancyfoot[CE]{\vspace{-7.5pt}\hspace{-13.5cm}\includegraphics{headers/RF.pdf}}
% \fancyfoot[RO]{\footnotesize{\sffamily{1--\pageref{LastPage} ~\textbar 
% \hspace{2pt}\thepage}}}
% \fancyfoot[LE]{\footnotesize{\sffamily{\thepage~\textbar\hspace{3.45cm}
% 1--\pageref{LastPage}}}}
% \fancyhead{}
% \renewcommand{\headrulewidth}{1pt} 
% \renewcommand{\footrulewidth}{1pt}
% \setlength{\arrayrulewidth}{1pt}
% \setlength{\columnsep}{6.5mm}
% \setlength\bibsep{1pt}

\twocolumn[
  \begin{@twocolumnfalse}
\noindent\LARGE{\textbf{Insights into the mechanisms of electromediated gene
delivery and application to the loading of giant vesicles with negatively
charged macromolecules$^\dag$}}
\vspace{0.6cm}

\noindent\large{\textbf{Thomas Portet,\textit{$^{a\ b}$} Cyril
Favard,\textit{$^{c}$} Justin Teissi\'e,\textit{$^{a}$} David S.
Dean,\textit{$^{b}$} and Marie-Pierre
Rols\textit{$^{a}$}$^{\ast}$}}\vspace{0.5cm}
%Please note that \ast indicates the corresponding author(s) but no footnote
%text is required. 

\noindent\textit{\small{\textbf{Received Xth XXXXXXXXXX 20XX, Accepted Xth
XXXXXXXXX 20XX\newline
First published on the web Xth XXXXXXXXXX 200X}}}

\noindent \textbf{\small{DOI: 10.1039/b000000x}}
\vspace{0.6cm}
%Please do not change this text.

\noindent \normalsize{We present experimental results regarding the
electrotransfer of plasmid DNA into phosphatidylcholine giant unilamellar
vesicles (GUVs). Our observations indicate that a direct entry is the
predominant mechanism of electrotransfer. A quantitative analysis of the DNA
concentration increments inside the GUVs is also performed, and we find that our
experimental data are very well described by a simple theoretical model in which
DNA entry is mostly driven by electrophoresis. Our theoretical framework allows
for the prediction of the amount of transfered DNA as a function of the electric
field parameters, and thus paves the way towards a novel method for
encapsulating with high efficiency not only DNA, but any negatively charged
macromolecule into GUVs.}

\vspace{0.5cm}
 \end{@twocolumnfalse}
  ]

%%%%%%%%%%%%%%%%%%%%%%%%%%%%%%%%%%%%%%%%%%%%%%%%%%%%%%%%%%%%%%%%%%%%%%%%%%%%%%%%
%%%
%%%%%%%%%%                              INTRO                           
%%%%%%%%%%
%%%%%%%%%%%%%%%%%%%%%%%%%%%%%%%%%%%%%%%%%%%%%%%%%%%%%%%%%%%%%%%%%%%%%%%%%%%%%%%%
%%%
\section{Introduction}
\label{sec:intro}
% Footnotes
\footnotetext{\dag~Electronic Supplementary Information (ESI) available: one
movie and associated legend. See DOI: 10.1039/b000000x/}
\footnotetext{\textit{$^{a}$~Institut de Pharmacologie et de Biologie
Structurale, CNRS UMR 5089, Universit\'e Paul Sabatier, Toulouse, France. }}
\footnotetext{\textit{$^{b}$~Laboratoire de Physique Th\'eorique, CNRS UMR 5152,
Universit\'e Paul Sabatier, Toulouse, France. }}
\footnotetext{\textit{$^{c}$~Institut Fresnel, CNRS UMR 6133, Universit\'es Aix-Marseille,
Marseille, France. }}
\footnotetext{\textit{$^{\ast}$~Tel: +33 5 61 17 58 11; E-mail: rols@ipbs.fr. }}

%% Nouvelle Intro

Lipid membranes are essential constituents of living organisms, especially
because of their impermeability to ions and hydrophilic molecules. This
impermeability allows for the crucial compartmentalization necessary for life to
develop \cite{Mou05}. However, the barrier presented by the membrane  turns out
to be a hurdle in many biotechnological applications such as gene delivery or
encapsulation of charged compounds into giant liposomes, \latin{i.e.} when one
needs to force the way of a charged molecule through a lipid bilayer.

As far as the loading of giant unilamellar vesicles (GUVs) is concerned, the
most common procedure is to directly prepare the liposomes in a medium
containing the desired compound. Fabrication of vesicles trapping
salt-containing solutions can be achieved \latin{e.g.} by hydration of a hybrid
film of lipids and agarose \cite{Hor09}. Also, natural swelling of GUVs
encapsulating high molecular mass DNA was shown to be possible in presence of
moderate concentrations of magnesium ions \cite{sat03}. Unfortunately, these
techniques still suffer from the drawbacks inherent to the gentle hydration
method: the resulting vesicles exhibit a wide size heterogeneity and their
unilamellarity can not be guaranteed. The well known electroformation protocol
\cite{Ang86} has a higher yield of unilamellar vesicles \cite{Rod05} . This
protocol  was recently refined in order to make electroformation possible  with
solutions containing various compounds, and in particular significant amounts of
ions. Performing the electroformation in a flow chamber allows the encapsulation
of high ionic strength solutions or large dextrans in the so formed GUVs
\cite{Est05}. However, this technique requires the use of a sophisticated setup
and does not seem easy to implement. Recently proposed modifications of the
electroformation protocol enabled the formation of GUVs under physiological
conditions \cite{Pot08,Mon07}. Nevertheless, these methods have not been tested
with large compounds of several tens of kDa, nor with highly charged molecules.
Another solution was proposed by Stachowiak \latin{et al.}, who designed a GUV
formation procedure using a pulsed microfluidic jet that deforms a planar
bilayer into a vesicle \cite{Sta08}. This technique does produce unilamellar
vesicles of homogeneous radii, and any solution could in principle be trapped in
the GUVs. Nevertheless, implementing  the microfluidic setup is not a trivial
task.

For the purpose of gene delivery to living cells, viral based  methods are
available; these are quite efficient but their safety has been questioned
\cite{Mar99}. Chemical methods relying on the formation of DNA complexes with
positively charged molecules are also a topic of active research \cite{Par03};
such methods, although safer, are less efficient than the viral methods. Among
physical methods, electropermeabilization is one of the most widely used.
Indeed, the cell membrane can be safely and transiently permeabilized in a very
elegant manner  by applying electric pulses \cite{neu72}.  Providing that  the
pulses are of sufficient duration \cite{rol98} and amplitude \cite{tei93},
otherwise non permeant molecules of therapeutic interest can enter the cytoplasm
of mammalian cells. This fact has led to two clinical applications:
electrochemotherapy \cite{bel93} and electrogenetherapy \cite{esc09}. The former
involves small molecules such as cisplatin or bleomycin, and the latter  larger
molecules such as plasmid DNA.

As was described in \cite{gol02}, the mechanisms of molecular uptake under
electropermeabilization, although poorly understood for the moment, depend
strongly on whether the transfered objects have molecular or macro-molecular
sizes. Whereas smaller ones seem to be able to freely cross electropermeabilized
cell membranes, macromolecules exhibit an intermediate interaction with the
membrane, the degree of this interaction being correlated with the ultimate
transfection efficacy. A precise description of the electropermeabilized
membrane would help to design safer and  more efficient protocols; this is a
major motivation for investigating the behaviour of simpler model systems. To
our knowledge, DNA electrotransfer into liposomes was first studied by
Chernomordik \latin{et al.} \cite{che90}. They found that high molecular mass
DNA could enter DPPC/cholesterol (7:3, mol:mol) LUVs, \latin{via}
endocytosis-like vesicles which shielded the electrotransferred  DNA from the
internal medium. The internalization mechanism they proposed was based on their
observation of the characteristic fluorescence of DNA/ethidium bromide (EB) complexes {after}
sonication of EB-loaded vesicles which were  pulsed in the presence of DNA.
However, they could not directly observe the liposomes because of their small
size. The conclusions of \cite{che90} were subsequently questionned by Lurquin
and Athanasiou \cite{lur00}, who observed that giant ($\approx 10\  \mu$m)
EB-loaded DPPC liposomes pulsed with DNA did actually show the bright
fluorescence of DNA/EB complexes {before} (or even without) sonication. These
results support a mechanism involving electropores and a direct entrance of DNA
into the liposomes. No endocytosis-like vesicles occur during this mechanism,
and this direct entrance allows the DNA to be immediately in contact with the
medium inside the liposomes. Although electropores could not be unequivocally
observed, these results are at variance with the findings of  \cite{che90}. 

In this paper we describe high molecular mass DNA electrotransfer experiments
performed on egg phosphatidylcholine (EggPC) GUVs. We present qualitative
observations regarding the pathway of electromediated DNA entry into the
liposomes, and quantitative results concerning the evolution of the inner DNA
concentration as a function of the number, amplitude and duration of the applied
electric pulses. By comparing our experimental data to a simple theoretical
model, we are able to address the two following key questions:
%\begin{enumerate}
\textit{(i)} what is the mechanism of DNA electrotransfer across pure lipid
membranes?
 and \textit{(ii)} is it possible to control/predict the amount of
electrotransfered DNA?
%\end{enumerate}
As well as being of interest for its comparison with the uptake of DNA by living
cells, this study presents a method for loading vesicles with negatively charged
macromolecules which may have a number of academic  and practical applications.

%%%%%%%%%%%%%%%%%%%%%%%%%%%%%%%%%%%%%%%%%%%%%%%%%%%%%%%%%%%%%%%%%%%%%%%%%%%%%%%%
%%%
%%%%%%%%%%                   MATERIALS AND METHODS                      
%%%%%%%%%%
%%%%%%%%%%%%%%%%%%%%%%%%%%%%%%%%%%%%%%%%%%%%%%%%%%%%%%%%%%%%%%%%%%%%%%%%%%%%%%%%
%%%

\section{Materials and methods}
\subsection{GUVs} 

Egg yolk L-$\alpha$-phosphatidylcholine (EggPC) and 
{L-$\alpha$-phosphatidylethanolamine-N-(lissamine rhodamine B sulfonyl)}
(Rhodamine PE) were purchased from Avanti Polar Lipids (Alabaster, AL). Lipids
were diluted in chloroform, at a mass concentration of 0.5 mg/mL, and stored at
-20\degre C. Rhodamine PE dye was added at a concentration of 1 mol \%.

The vesicles were prepared at room temperature using the electroformation
protocol \cite{Ang86}, in an aqueous solution of 240~mM sucrose (internal
solution) and subsequently diluted in an aqueous solution of 260~mM glucose and
1~mM sodium chloride (external solution). The pH of the external solution was
adjusted to 7 using a 1~mM phosphate buffer (KH$_2$PO$_4$/K$_2$HPO$_4$, Merck,
Darmstadt, Germany). The pH of the internal solution was measured to be 6.6.
This slight initial pH asymmetry did not have any effect on our experiments:
non-pulsed vesicles were stable for several tens of minutes, and no increase in
DNA concentration inside a non-pulsed GUV could be detected. Furthermore, this
initial pH difference will vanish after a few permeabilizing pulses because of
the mixing of inner and outer solutions.  The conductivities of the internal and
external solutions were measured with conductivitymeter HI 8820  (Hanna
Instruments, Lingolsheim, France), and had values of approximately  $1.5\times
10^{-3}$~S/m and $4.5\times 10^{-2}$~S/m, respectively. The osmolarities,
measured with Osmomat 030 osmometer (Gonotec, Berlin, Germany), were
approximately 280 mOsm/kg and 300 mOsm/kg, respectively. These solutions are
typically used when working with GUVs for several reasons. The similar
osmolarities ensure that the GUVs have very little initial membrane tension, and the
low conductivities prevent Joule heating in the pulsation chamber. The sugar
asymmetry yields a density difference that allows sedimentation of the vesicles
to the bottom of the chamber and facilitates their localization.

Electroformation was performed as follows. A small volume (15 $\mu$L) of the
lipid solution in chloroform was deposited on the conducting sides of glass
slides coated with indium tin oxide. The glasses were then kept for two hours
under vacuum in a desiccator to remove all traces of the organic solvent.
Afterwards, the plates, spaced by a 1 mm thick silicon frame (Electron
Microscopy Sciences, Hatfield, PA) were assembled into a chamber. The chamber
was filled with the sucrose solution (internal medium). The slides were
connected to an AC field function generator (AC Exact, model 128; Hillsboro, OR)
and sinusoidal voltage of 25 mV peak to peak at 10 Hz was applied. The voltage
was increased by 100 mV steps every 5 minutes, up to a value of 1225 mV and
maintained under these conditions overnight. Finally, square-wave AC field of
the same amplitude was applied at 5 Hz for one hour in order to detach the GUVs
from the slides.

\subsection{DNA} 
A 4.7 kbp plasmid (MW $\sim 3.10^6$~Da) pEGFP-C1 (Clontech, Mountain View, CA)
carrying the green fluorescent protein  gene controlled by the
cytomegalovirus promoter was stained stoichiometrically with the DNA
intercalating dye TOTO-1 (Molecular Probes, Eugene, OR). Staining was carried
out at a DNA concentration of 1 $\mu$g/$\mu$L for 60 minutes on ice, at a
basepair to dye ratio of 20. Plasmids were prepared from Escherichia coli
transfected cells by using Maxiprep DNA purification system (Qiagen, Chatsworth,
CA). The plasmid was then diluted in the pulsation buffer at a mass
concentration of 1 $\mu$g per 100 $\mu$L, a value typically used in cell
electrotransfection experiments.  
This plasmid was chosen because of its large size similar to those of the DNAs
used in \cite{che90,lur00}, and also because it is classically used in gene
electrotransfer studies where it was observed to form discrete interaction sites
at the cell surface during electropermeabilization experiments \cite{gol02}.

\subsection{Electropulsation}
Pulsation chambers were similar to those used in \cite{por09}.  Two parallel
copper strips {(3M, Cergy-Pontoise, France)} were stuck on a glass slide 0.5 cm
apart. A glass coverslip was then stuck onto the glass slide with heated
parafilm. The cavity between the slide and the coverslip was filled with 30
$\mu$L of the buffered glucose solution and 2 $\mu$L of the GUV solution. 
Electropulsation was performed directly on the microscope stage, by applying a
series of 10, 20 or 40 pulses of 0.5, 1 or 5 ms duration at 0.33 Hz repetition
frequency with a $\beta$-Tech pulse generator ($\beta$-Tech, L'Union, France).
Field amplitudes $E_0$ were adjusted according to the initial vesicle radius
$R_0$ in order to impose initial induced transmembrane voltages
$\Delta\psi_0=(3/2)R_0E_0$ \cite{Neu89} ranging between 0.5 and 2 V, and as a
consequence varied between 20 and 80 kV/m from one experiment to another.

\subsection{Confocal microscopy}
All experiments were performed under an inverted confocal microscope (Zeiss
LSM510; Carl Zeiss, Jena, Germany) equipped with a 63$\times$ Zeiss objective
for fluorescence imaging. For the red channel (membrane labeled with Rhodamine),
excitation at 543 nm was provided by a HeNe laser, and emission filter was a 560
nm long pass. For the green channel (DNA labeled with TOTO-1), excitation at 488
nm was provided by a Ar laser, and emission filter was a 500-530 nm band pass.
Images were acquired sequentially. Acquisition time was 2~s  (1~s for each
channel), thus one image on each channel could be acquired between every two
consecutive pulses, separated by 3~s. The characteristic time for homogenizing
the concentration of a molecule of diffusion coefficient $D$ within an object of
size $l$ is on the order of $l^2/D$. For plasmid DNA of diffusion coefficient
$D=10^{-12}$~m$^2$s$^{-1}$\cite{luk00} and regions of inhomogeneous DNA
concentration with sizes in the micrometer range, this yields characteristic
times on the order of a few seconds. Thus in our experiments, DNA concentrations
had enough time to homogenize between consecutive pulses. The use of confocal
microscopy for these experiments is crucial, as it ensures that the measured
fluorescence intensities are simply proportional to the local concentration of
the fluorescent molecules \cite{Li10}.

\subsection{Image and data processing}

We were interested in two different quantities, both measurable via confocal
fluorescence microscopy: the vesicle radius $R$ and the ratio of the DNA
concentrations inside and outside the vesicle, $c/c_0$. As we work in a dilute
water solutions with similar physicochemical properties, the concentration of a
fluorescently-labeled molecule is simply proportional to the mean fluorescence
intensity due to this molecule, thus the quantity $c/c_0$ is equal to $I/I_0$,
the ratio of the mean fluorescence intensities inside and outside the vesicle,
which is easily measurable in our experiments.

Our images were composed of two channels: a red one for Rhodamine PE (lipid
membrane) and a green one for TOTO-1 (DNA). GUVs size measurements were
performed on the red channel in a semi-automatic manner via custom-written
programs, and relative DNA amount quantification was performed  on the green
channel automatically, by computing the mean fluorescence intensity {inside} the
vesicle within a disk whose radius was chosen at a value of 80 \% of the
measured radius of the vesicle (we checked that no significant differences were
seen for disk size values ranging from 60 to 90 \% of the GUV size). This
procedure was repeated for each image, so we were able to plot the vesicle size
and the DNA quantity trapped in the vesicle as a function of time. We also
computed the mean fluorescence intensity {outside} the vesicle. It was checked
that this value did not change significantly after each pulse, and thus we used
the value computed from the first image.

All images had non-zero mean fluorescence intensity values, even inside a
non-pulsed vesicle where no DNA was present, or in pictures of pulsation
chambers filled with pure water. This was due to the bias level of the confocal
microscope sensors. In order to properly quantify DNA concentrations, we
estimated this bias fluorescence level inside a non-permeabilized GUV containing
no DNA, and substracted this quantity from the measured internal and external
DNA fluorescence intensities to obtain the fluorescence intensity ratio $I/I_0$
equal to the concentration ratio $c/c_0$.

Data for $c/c_0$ exhibited rather strong fluctuations and were smoothed using a
$n=3$ moving average technique. Error bars in Fig.~\ref{fig:fluo} were obtained
by computing the standard deviation of the fluctuations around the constant
value of $c/c_0$ after the application of pulses, for each experiment. Image and
data processing tasks were performed with Matlab (The Mathworks, Natick, MA) 
and ImageJ (National Institute of Health, Bethesda, MD).

\subsection{Numerical computation of the electric field near a permeabilized
vesicle}
\label{subsec:matmetnum}

As described in the theoretical section of this paper, we were interested in
flow of DNA into 
a permeabilized vesicle. In the limit where the electrophoretic force on a DNA
molecule
dominates the osmotic force (due to concentration gradients), which we argue is
the case
in our experiments, the entry of DNA into the cell can be computed from the flux
of the electric field across the cathode-facing permeabilized region of the GUV.
 The permeabilized region is defined as a spherical cap of angle $\theta$ facing
the cathode, and we also assume that an identical region, similarly
permeabilized to ionic currents,  exists on the anode-facing side.

We computed numerically this flux $\phi$ as a function of $\theta$ using the
finite element calculus software Comsol Multiphysics (Comsol, Burlington, MA). 

The GUV was modeled as a spherical shell of radius $R$ and internal electrical 
conductivity $\sigma_i$, embedded in an aqueous solution of electrical
conductivity $\sigma_e$.  
Laplace's equation for the electric potential $\psi$ \cite{Jac75}
\begin{equation}
\nabla \cdot \sigma \nabla \psi =0
\end{equation} 
was solved numerically in a cylindrical box of radius $L$ and heigth $2L$,
taking advantage of the axial symmetry of this problem. An electric field of
amplitude $E_0=V_0/L$ was applied in the z-direction parallel to the cylinder
axis, by imposing the Dirichlet boundary conditions $\psi=V_0$ and $\psi=-V_0$
on the top and the bottom faces of the cylinder, hence their names of anode and
cathode, respectively.
The membrane was modeled using thin layer boundary condition with a thickness
$d=4$~nm and a membrane conductivity $\sigma_m=5\times 10^{-7}$~S/m
\cite{Puc06}. For representing the permeabilized regions of the vesicle facing
the cathode and anode, we replaced the membrane conductivity $\sigma_m$ by the
external conductivity $\sigma_e$ on a spherical cap of angle $\theta$. Numerical
integration of $\mathbf{E} = - \nabla \psi$ on that surface thus gave us the
value of the flux $\phi$ as a function of $\theta$. 

Different values of the model parameters $R$, $E_0$, $\sigma_i$ and $\sigma_e$
corresponding to realistic experimental conditions were explored, and are given
in the theoretical section and in Figs.~\ref{fig:flux}~and~\ref{fig:flux_2}.
Typical values were $R=10^{-5}$~m, $E_0=50$ kV/m, $\sigma_i=10^{-2}$~S/m and
$\sigma_e=4\times 10^{-2}$~S/m.

\subsection{Determination of the relative concentration increments $\delta x$}

The amount of DNA transfered into a GUV per pulse can be considered in terms of
a relative concentration increment $\delta x$, \latin{i.e.} the difference
between the value of the DNA concentration inside a liposome after a pulse and
its value  before the pulse, normalized by the outer DNA concentration. This
quantity was measured experimentally by substracting the value of the mean
fluorescence intensity ratio $I/I_0$ before the pulse to the value of $I/I_0$
after the pulse.

Within the framework of our model, $\delta x$ can also be predicted using
Eq.~\ref{eqn:deltax}, where the electric pulse parameters $E_0$ and $\tau$ are
chosen by the experimenter, the vesicle size $R$ is easily measured, and the DNA
electrophoretic mobility is $\mu=-3.75\times 10^{-8}$~m$^{2}$V$^{-1}$s$^{-1}$
\cite{Ste97}. The value of the flux factor $f'(0)\theta$ is estimated to be on
the order of 0.2 (see section~\ref{sec:theo} and Fig.~\ref{fig:flux_2}).

%%%%%%%%%%%%%%%%%%%%%%%%%%%%%%%%%%%%%%%%%%%%%%%%%%%%%%%%%%%%%%%%%%%%%%%%%%%%%%%%
%%%
%%%%%%%%%%                              RESULTS                         
%%%%%%%%%%
%%%%%%%%%%%%%%%%%%%%%%%%%%%%%%%%%%%%%%%%%%%%%%%%%%%%%%%%%%%%%%%%%%%%%%%%%%%%%%%%
%%%

\section{Experimental results}

% petite intro:
% 		combien de vesicules,
%		parametres electriques, 
We performed DNA electrotransfer experiments with 21 EggPC liposomes. Either 10,
20 or 40 pulses were applied at 0.33 Hz. 
 The durations of the applied pulses were 0.5, 1 or 5 ms, and correspond to
those commonly used for electromediated gene transfer \cite{rol98}. The longest
pulse duration $\tau$ was observed to maximize the amount of transfered DNA, so
most of the experiments were done at this value of 5 ms. Depending on the
experiment, the pulse amplitude varied from 20 to 80 kV/m.  The applied field
magnitude $E_0$ was tuned in order to obtain initial induced transmembrane
voltages (at the poles facing the cathode and the anode) ranging between 0.5 and
2 V. This was estimated from
the formula for the voltage drop at a point on the membrane whose radial vector
(of magnitude
$R$) makes an angle $\theta$ with the direction of the applied
field\cite{Neu89}:
\begin{equation}
\Delta \psi = -\frac{3}{2}E_0R\cos(\theta).
\end{equation}
This formula is valid when the conductivity of the membrane is much smaller than
those of the internal and external solutions  and when the thickness of the
membrane $d$ is small compared to the vesicle size $R$, which is true here.
 In some cases we were able to keep several vesicles in the optical field, thus
we could gather data from several liposomes during the same experiment.
Typically, we tuned the field magnitude according to the size of the largest GUV
we could see. In \cite{por09} it was established for EggPC and DOPC liposomes
that there was a critical value of the absolute value of the  transmembrane
voltage $\Delta \psi_c$ below which vesicles were not permeabilized and showed
no change under the application of the electric
field. Sufficiently large vesicles were seen to eject lipid material in the form
of small vesicles and tubules, however this lipid ejection caused the vesicle
radius to diminish and at constant field,
when the radius $R$  fell below a threshold value the vesicle was no longer
visibly affected by the field.
In agreement with the results of \cite{por09} it was found that the smallest
liposomes were not permeabilized and no DNA fluorescence could be detected
inside them. For vesicles of intermediate sizes, the permeabilized area is
smaller than that of the largest  GUV on which the field of view was centered.
Consequently these vesicles of intermediate sizes should be expected to exhibit
relatively  low amounts of  DNA uptake. However, we kept such vesicles in our
analysis, and coherent results for the increments of the internal DNA
concentration were also obtained with them (section~\ref{sec:theo}). 

\begin{figure*}%[!h]
\begin{center}
\centerline{\includegraphics[width=16cm]{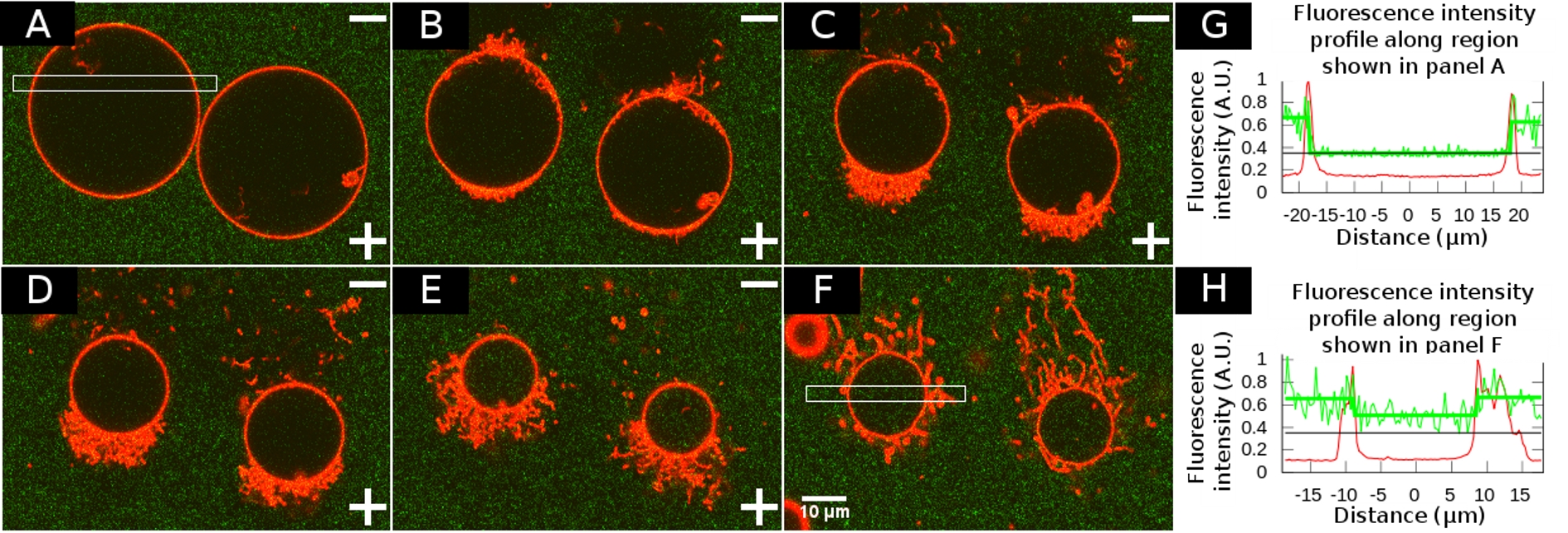}}
\caption{\textbf{A-F:} Typical raw images of one experiment. Rhodamine PE (lipid
membrane) and TOTO-1 (DNA) are shown in red and green respectively. 20 pulses of
39 kV/m amplitude and 5 ms duration were applied at 0.33 Hz (one pulse every
three seconds, and a total pulsation time of 60 s). Images were taken at 0, 6,
12, 18, 54 and 222 s; time origin was defined as the onset of the acquisition,
and the first pulse was applied after about 2.5 s. Vesicles are thus shown after
having received 0, 2, 4, 6, 18 and 20 electric pulses, respectively.  Field
polarity is indicated on pictures acquired during electropulsation (+ and -
electrodes). The last picture (panel F) was acquired more than two minutes after
the pulse train, and shows the stability of the vesicles loaded with DNA.
Tubular structures associated with vesicle size decrease and similar to those
described in \cite{por09} are present.   A scalebar of 10 $\mu$m length can be
seen on panel F. \textbf{G and H:} Fluorescence intensity profiles along white
rectangular regions shown in panels A and F, respectively. Thick green line
represents the mean value of TOTO-1 fluorescence intensity inside, on the left,
and on the right of the GUV. Thick black line represents the bias fluorescence 
level, a non zero fluorescence value corresponding in reality to a zero DNA
concentration. This value was thus substracted for subsequent data
analysis.}\label{fig:photo}
\end{center}
\end{figure*}

% mot sur les mecanismes plus photo
As can be seen in Fig.~\ref{fig:photo}, it is possible to optically detect DNA
entering the vesicle - a movie is also provided as ESI\dag \ on the journal
website. The images show vesicles exhibiting lipid loss via formation of tubular
structures.  These structures were created on the anode facing side of the GUV
and appeared to remain attached to the vesicle and stable over a few minutes, as
 previously reported in \cite{por09} for DOPC liposomes. 

Darker zones denoting a local depletion of DNA from the outer region of the
vesicle facing the cathode were observed in a small number of experiments, as
shown in Figs.~\ref{fig:photo}B and C. This phenomenon was quite rare, and could
only be detected following the first few electric pulses.  It may be due to
hydrodynamic flows resulting from the leakage of the inner sucrose solution or
from osmotic effects. However, this local depletion did not have any appreciable
influence on the subsequent behavior of the vesicle around which it was seen.

We confirm the results of Lurquin and Athanasiou \cite{lur00}, namely the free
entrance of DNA through pores formed by the electric field. Indeed if an uptake
via endocytosis-like vesicles was the predominant mechanism, we would detect the
presence of such objects in most of our experiments. Even though presumably
quite small (some tens or hundreds of~nm), some of these endocytosis-like
vesicles should be created in the focal plane where the transmembrane voltage
induced by the electric field is maximal, and should thus appear in our images. 
This was not the case. Therefore our pictures clearly speak in favour of an
entrance via pores. We  mention that we sometimes noticed  endocytosis events,
but they were very rare, and these phenomena always took place {\em after}
application of the electric pulses. A detailed explanation of this  rare uptake
via vesicles  is beyond the scope of the present work. However, we stress that
an explanation based on the mechanism described in \cite{ang99} where long DNA
fragments of a few~kbp absorb on a patch of the GUV membrane before it bends
and pinches off toward the interior of the GUV  would not be satisfactory
because in that study the authors reported no DNA internalization within
liposomes made just of zwitterionic lipids, as our EggPC GUVs. 
 
% mot sur DOPC
%We further add that similar qualitative behavior  was also observed during
%preliminary experiments performed on DOPC vesicles, but the analysis and images
%presented in this paper were obtained with EggPC GUVs only.

\begin{figure*}%[htbp]
\begin{center}
\centerline{\includegraphics[width=16cm]{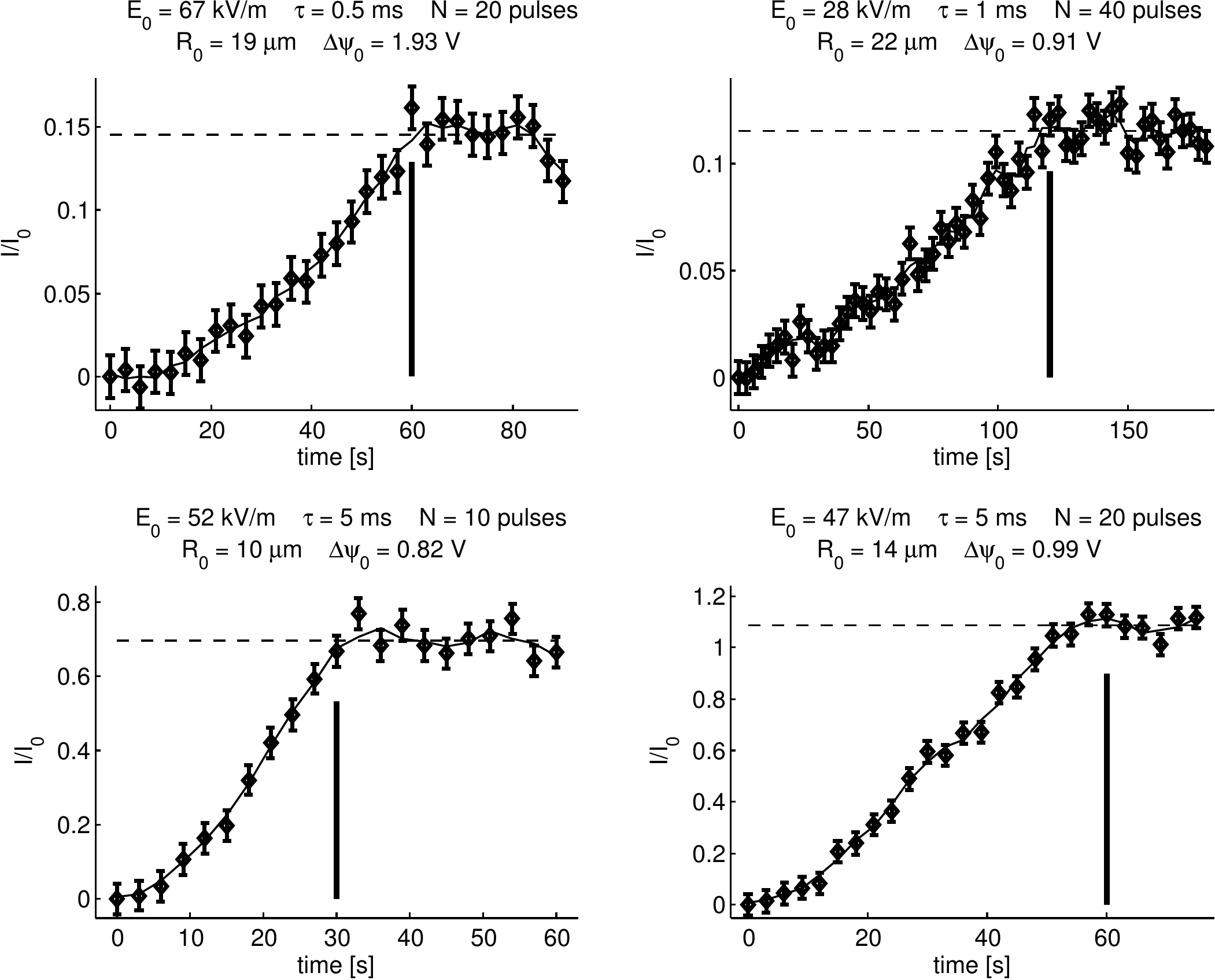}}
\caption{Plots of the relative mean fluorescence intensity inside the vesicle
$I/I_0$ (or equivalently the relative concentration of DNA $c/c_0$) as a
function of time, for 4 typical experiments. Diamond marks represent raw data,
and thin full lines these same data smoothed by a $n=3$ moving average
procedure. This smoothed values were used to compute the individual
concentration increments $\delta x$.  Error bars were obtained by computing the
standard deviation for the $I/I_0$ values after the pulse sequence. Final
relative concentration, obtained by averaging the $I/I_0$ values after the pulse
sequence, is shown with horizontal dashed lines. Thick vertical lines indicate
the end of the pulse sequence. Note the different horizontal and vertical scales
for each graph. Electric field parameters, initial vesicle radius $R_0$ and
initial induced transmembrane voltage $\Delta\psi_0$ are given on each panel.
One pulse was applied between every two consecutive data points (0.33 Hz
repetition frequency).}\label{fig:fluo}
\end{center}
\end{figure*}

% montrer les donnees experimentales
Typical examples of fluorescence experimental data, obtained as described in the
Materials and Methods section, are shown in Fig.~\ref{fig:fluo} (diamond marks).
These data for $I/I_0$ are the ratio of the mean fluorescence inside and outside
the GUV, and are equivalent to the relative DNA concentration $c/c_0$, where $c$
is the DNA concentration inside the liposome and $c_0$ the DNA concentration
outside the liposome. A ratio $I/I_0$ equal to 1 thus means that the DNA
concentrations inside and outside the vesicle are the same.  The field amplitude
$E_0$, the pulse duration $\tau$, the number of pulses $N$, the initial vesicle
radius $R_0$ and the initial transmembrane voltage $\Delta\psi_0$ induced by the
electric field are given in each panel. For each experiment, we observe an
increase of the amount of DNA inside the vesicle during the train of pulses,
along with a stabilization after the end of the pulse sequence. This clearly
shows that DNA can enter GUVs and remain trapped in a stable manner, without
significant leakage. We draw the reader's attention to the different scales used
on each panel: the final DNA concentrations inside the liposomes presented in
the two top panels  were much lower than those inside the liposomes subjected
to electric pulses of the longest duration, which are presented in the bottom panels. 

% parler de tous les controles (TOTO seul, arreter de pulser et voir que ea
%reste stable, puis reprise, voir que ca ne monte pas quand on ne pulse pas)
A number of  control experiments were performed. First, we checked that
fluorescence intensities inside the liposomes did not change if we did not apply
pulses. This verification was performed on long timescales (several minutes). As
a second control experiment, we applied a series of ten pulses and checked that
the fluorescence level, after having increased, remained stable until the
application of another train consisting of ten pulses of slightly higher
amplitude,  which induced a subsequent augmentation of the relative fluorescence
intensity. This is important as it proves that the DNA uptake is
induced directly by the electric field, ruling out the possibility that  the
vesicle remains permeable to DNA for some time after the field is cut and
emphasizing the importance of the electrophoretic force on the DNA as well as
the permeabilizing effect of the electric field on the membrane.
 We also carried out control experiments with  TOTO-1 but with no DNA, in order
to check that it was not unbounded dye we observed. The quantum yield of TOTO-1
is dramatically increased when bound to DNA. As a consequence, images acquired
with same microscope settings and in the absence of DNA  should be much less
fluorescent than those with DNA. This was indeed the case (in fact the signal
from TOTO-1 alone was too weak to be detected), thus proving that it was indeed
fluorescently labeled DNA we were monitoring.

% c/c0>1
Interestingly, we found that sometimes the final DNA concentration inside the
GUV could exceed the outer DNA concentration ($c/c_0=I/I_0>1$). That can be seen
for example on the bottom right panel of Fig.~\ref{fig:fluo}. This is quite
surprising, and means that if the DNA enters the vesicle from the cathode facing
pole driven by the electric field, it is not able to freely leak out the vesicle
from the other pole. 
Indeed, if the permeabilization structures were the same on each hemisphere of
the liposome, the  maximal DNA concentration ratio possible should be $c/c_0=1$.
This is not the case in our experiments. This observation, along with the
asymmetric formation of tubules, highlights the different membrane
reorganization and symmetry breaking between the two hemispheres.
The possibility of exceeding the external concentration of DNA within the
vesicle also implies that the
electrophoretic force due to the applied field overcomes the osmotic force due
to the difference in the
concentrations between the inside and outside of the vesicle, this will be
discussed in the theory section
which follows.   
% size decrease et fits pour 5 ms
We were also able to monitor the size decrease of the GUVs during the
experiments. Applying the method previously described in \cite{por09}, we
obtained the fraction of permeabilized area lost per pulse $\lambda$ and the
critical transmembrane potential difference $\Delta\psi_c$ required to observe
the size decrease of the GUVs. For the pulses of 5 ms duration (the same
duration $\tau$ as used in \cite{por09}), we find by averaging on all our
experiments: $\lambda \approx 0.32$ and $\Delta\psi_c\approx 0.50$ V. The value
found for $\lambda$ agrees well with the previously reported value of 0.31,
which means that the GUVs still lose $\sim$ 30 \% of their permeabilized area
(area where $\Delta\psi_c$ is exceeded) per pulse \cite{por09}. However, the
$\Delta\psi_c$ value of 0.50 V found  here is  much smaller than the value 0.89
V found for vesicles pulsed in the absence of DNA. This suggests that the
presence of DNA in the external medium, and/or inside the vesicle after a few
permeabilizing pulses, tends to destabilize the vesicle membrane and makes it
easier to electropermeabilize, but does not affect the fraction of permeabilized
area expelled from the liposome.

%%%%%%%%%%%%%%%%%%%%%%%%%%%%%%%%%%%%%%%%%%%%%%%%%%%%%%%%%%%%%%%%%%%%%%%%%%%%%%%%
%%%
%%%%%%%%%%                              THEORY                          
%%%%%%%%%%
%%%%%%%%%%%%%%%%%%%%%%%%%%%%%%%%%%%%%%%%%%%%%%%%%%%%%%%%%%%%%%%%%%%%%%%%%%%%%%%%
%%%

\section{Theoretical analysis of the relative DNA concentration increments}
\label{sec:theo}

The theoretical interpretation of our experimental results is rather difficult
as we do not have 
explicit access to the way in which the vesicles are permeabilized.  

In this section we will present a simple model describing the relative DNA
concentration increment $\delta x=\delta c / c_0$ inside a vesicle of radius R,
caused by an electric pulse of amplitude $E_0$ and duration $\tau$. The DNA
concentration increment inside the vesicle, per pulse, is denoted by $\delta c$,
and the constant outside concentration by $c_0$. We will assume that the major
contribution to $\delta x$ is due to the pulling of the DNA molecules by the
electric field, and we will see that our 360 experimental data points agree with
this hypothesis. 

The DNA concentration $c$ is governed by an electrodiffusion equation
\begin{equation}
\frac{\partial c }{\partial t} = - \nabla \cdot \mathbf{j} \ ,
\label{eqn:dcdt}
\end{equation}
where $\mathbf{j}$ is the thermodynamic current $\mathbf{j} = -D\nabla c + \mu c
\mathbf{E}$. The first term is a diffusion term that depends on the molecule
diffusion coefficient $D$. The second term is an electrophoretic term depending
on the local field $\mathbf{E}$ and the electrophoretic mobility of the molecule
$\mu$ (the steady state velocity $\mathbf{v}$ of the molecule in a field
$\mathbf{E}$ is given by $\mathbf{v}=\mu\mathbf{E}$).
The DNA concentration increment $\delta c (t)$  inside an object $\mathcal{V}$
of volume $V$ during a time $t$ is given by
\begin{equation}
\delta c (t)= \int_0^t \left( \frac{1}{V} \int_{\mathcal{V}} \frac{\partial c
}{\partial t} \ d\mathbf{x} \right) \ dt \ ,
\end{equation}
which can be written using Gauss' theorem:
\begin{equation}
\delta c (t)= \int_0^t \left( \frac{1}{V} \int_{\mathcal{S}} -\mathbf{j} \
d\mathbf{S} \right) \ dt \ . 
\end{equation}
We easily estimate the relative importance of electrophoretic forces and DNA
diffusion 
by comparing the diffusive or osmotic part of the current per concentration
${\bf r}_{os}={\bf j}_{os}/c=-D\nabla \ln(c)$ and the electrophoretic part ${\bf
r}_{ep}={\bf j}_{ep}/c =  \mu {\bf E}$. 
We use the following values of DNA in solution: $\mu=-3.75\times
10^{-8}$~m$^{2}$V$^{-1}$s$^{-1}$ \cite{Ste97} and $D=10^{-12}$~m$^2$s$^{-1}$
\cite{luk00}. We use the  estimations
$|{\bf r}_{os}| \sim D|\ln(c/c_0)|/R$ (where $c$ is the average DNA
concentration in the vesicle and $R$ the vesicle radius) and $|{\bf r}_{ep}|\sim
|\mu| E_0$, this gives
\begin{equation}
\frac{|{\bf r}_{os}|}{|{\bf r}_{ep}|}\sim \frac{ D|\ln(c/c_0)|}{|\mu|E_0 R}\sim
{10^{-4}\ln\left(\frac{c}{c_0}\right)}
\end{equation}
for the lower range of permeabilizing applied voltages ($E_0 R \sim 1$~V, cf
subsection~\ref{subsec:matmetnum}). We thus see that the effects of diffusion
are
negligible except for perhaps the first pulse when $c\sim 0$.  

As far as the ionic currents are concerned we assume that the vesicle is
permeablized in a symmetric way at both the cathode and anode facing sides. The
permeabilized region subtends an angle 
$\theta$ in the direction of the applied field. As mentioned in the Material and
methods section we model the permeabilized region by replacing the membrane
conductivity by that of the external solution; this 
is clearly a simplification. As far as the transport of DNA is concerned the
fact that macropores are
only ever seen at the cathode facing pole suggests that although the conductive
properties of the
permeabilized regions at the anode and cathode facing poles are the same (i.e.
the transport properties
of these regions to small ions is the same), DNA can pass more freely at the
cathode facing side than 
the anode facing side. Thus in our analysis we posit that DNA can be freely
electrophoretically conducted through the cathode facing permeabilized region.
In our model this is equivalent to saying that electrophoretic mobility of  DNA
in the membrane region is zero except in the permeabilized region
facing the cathode, where it takes its free solution value. The concentration
increment $\delta c (t)$ is simply related to the integral over time of the flux
of the thermodynamic current $\mathbf{j}$ across the surface $\mathcal{S}$. Let
us now consider a spherical object with fixed volume $V$ and let us compute the
concentration increment $\delta c$ from the onset of one pulse to the onset of
the subsequent pulse. Using a constant volume is a reasonable approximation
because in our experiments the average volume decrease caused by a single pulse
is still small compared to the initial volume of the vesicle.

\begin{figure}%[htbp]
%\begin{center}
\centerline{\includegraphics[width=8cm]{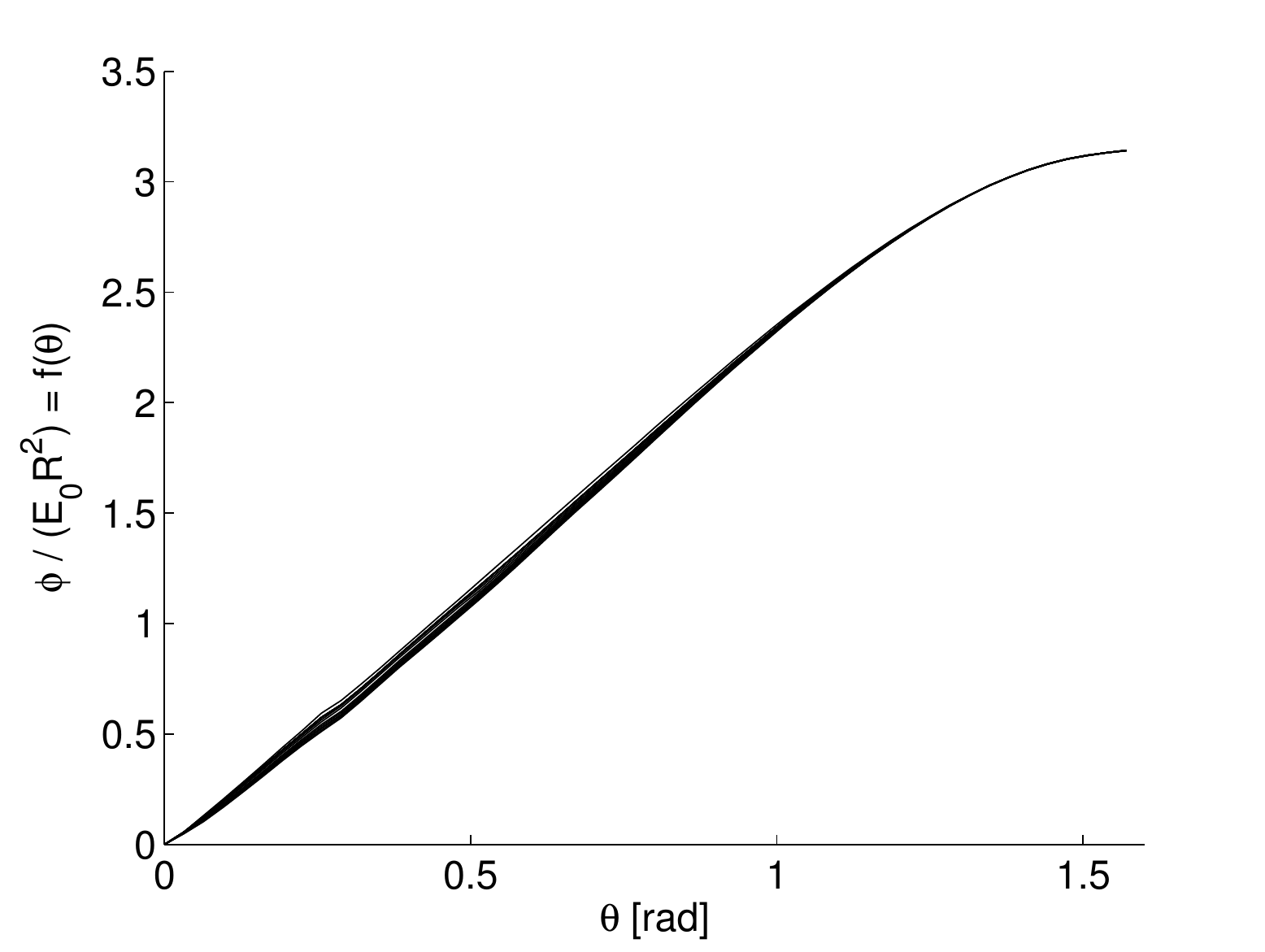}}
\caption{Flux $\phi$ of the electric field across a permeabilized spherical cap
of angle $\theta$. Flux values were computed numerically for different vesicle
radii $R$ ranging from 5 to 50 $\mu$m, and various electric field amplitudes
$E_0$ ranging from 10 to 500 kV/m. Internal and external conductivities
$\sigma_i$ and $\sigma_e$ were both equal to $4\times 10^{-2}$~S/m. Slight
deviations occur for curves corresponding to the different radii, and are due to
the different meshes used. All curves for the different field amplitudes exactly
superpose.}\label{fig:flux}
%\end{center}
\end{figure}

We will base the analysis of our experimental results on the following
assumptions (a) the dominant
transport process is electrophoretic and we neglect DNA diffusion (b) we assume
that over the period 
of the pulse a conducting hole is created at both poles of  membrane each having
 radius $a$. Concretely this means that the average radius  of the conducting
hole in the membrane averaged over the time of the pulse application is $a$.
Beyond the time averaging, this  is  a crude simplification, as the surface
permeable to ions may be larger than the one permeable to DNA, this is because
we cannot be sure that there is really a single macropore in the lipid bilayer,
instead of several smaller pores, or other types of defects. Nevertheless, we
expect the electroconvected flux of DNA into the vesicle to be of the same
order. Furthermore we will assume that DNA is not able to exit the vesicle at
the conducting region 
facing the cathode. 
Adopting the hypotheses above, upon a pulse of duration $\tau$ the increase in
concentration is given by
\begin{equation}
\delta c = \frac{|\mu| c_0 \tau}{V} \int_{\mathcal{S}_p} \textbf{E} \cdot
d\mathbf{S} \ .
\end{equation}
{\em i.e.} concentration increment $\delta c$ is just proportional to the flux
$\phi$ of the local electric field $\mathbf{E}$ across  the permeabilized
surface $\mathcal{S}_p$, the hole of radius $a$. In terms of the relative
concentration increments $\delta x=\delta c/c_0$, which are exactly the
differences between every two consecutive points of our experimental data for
$I/I_0$, this can be written
\begin{equation}
\delta x = \frac{3 |\mu| \tau}{4 \pi R^3} \ \phi \ .
\end{equation}
In order to be able to evaluate $\delta x$, we should be able to evaluate
$\phi$. Recall we assume that the permeabilized area is a spherical cap of angle
$\theta$ facing the cathode, that both DNA and ions can cross. Despite this
simplification of the transport problem no analytical solution to this problem
exists.  However, on purely dimensional grounds (assuming the limit of zero
membrane thickness), we can write 
\begin{equation}
\phi = E_0 R^2 f(a/R) \ ,
\label{eqn:phi}
\end{equation}
where $a$ is the pore radius such that $a=R\sin(\theta)$, and $f$ an unknown
function of the ratio $a/R$, or in other words a function of the permeabilized
angle $\theta$.  Note that in the limit where
the pore length is much greater than the pore radius, which is the case for
nanopores, Eq.~\ref{eqn:phi} will not be valid \cite{Wan10} and will depend explicitly on the
pore length.  
To check the validity of Eq.~\ref{eqn:phi}, we computed numerically $\phi$ for different field intensities
$E_0$, different vesicles radii R, and different angles $\theta$, as described
in the Materials and methods section. These results are presented in
Fig.~\ref{fig:flux}, where one can see that all the flux values superpose. These
plots are in fact plots of the function $f(a/R)=f(\theta)$, and show that
Eq.~\ref{eqn:phi} is valid. For a small angle $\theta$, we can make the linear
approximation that $f(\theta)\approx f'(0) \theta$, where $f'(0)$ is the slope
of the straight line giving $f(\theta)$. We finally obtain for the relative
concentration increments 
\begin{equation}
\delta x = \frac{3 |\mu| E_0 \tau}{4 \pi R} \ f'(0) \theta \ .
\label{eqn:deltax}
\end{equation}

\begin{figure}%[htbp]
%\begin{center}
\centerline{\includegraphics[width=8cm]{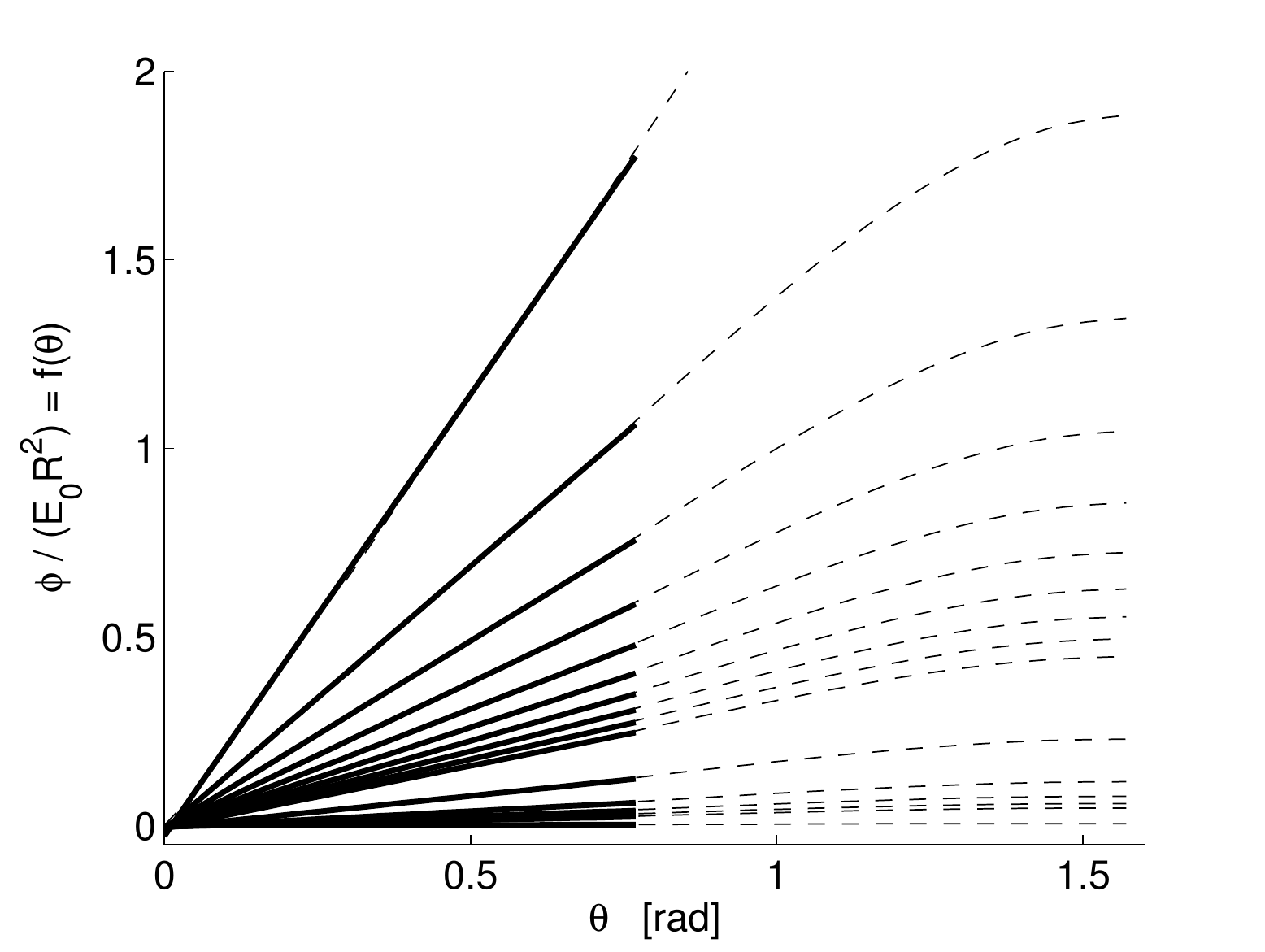}}
\caption{Flux $\phi$ of the electric field across a permeabilized spherical cap
of angle $\theta$. Flux values (dashed lines) were computed numerically for
different internal conductivities $\sigma_i$ of values $\sigma_e$, $\sigma_e/2$,
$\sigma_e/3$ ... $\sigma_e/10$, $\sigma_e/20$, $\sigma_e/40$ ... $\sigma_e/100$
and $\sigma_e/1000$ (from top curve to bottom curve). Vesicle radius $R$ was 10
$\mu$m, field amplitude $E_0$ was 60 kV/m, and external conductivity $\sigma_e$
was $4\times 10^{-2}$~S/m. Thick full lines are linear fits to numerical data.
These fits show that the linear approximation on $f(\theta)$ made to obtain
Eq.~\ref{eqn:deltax} is valid, and that for reasonable angles of the
permeabilized area $\theta$ and for internal and external conductivities of the
same order, $f'(0)\theta$ is of the order of~0.1--0.5.}\label{fig:flux_2}
%\end{center}
\end{figure}

\begin{figure}%[htbp]
%\begin{center}
\centerline{\includegraphics[width=8cm]{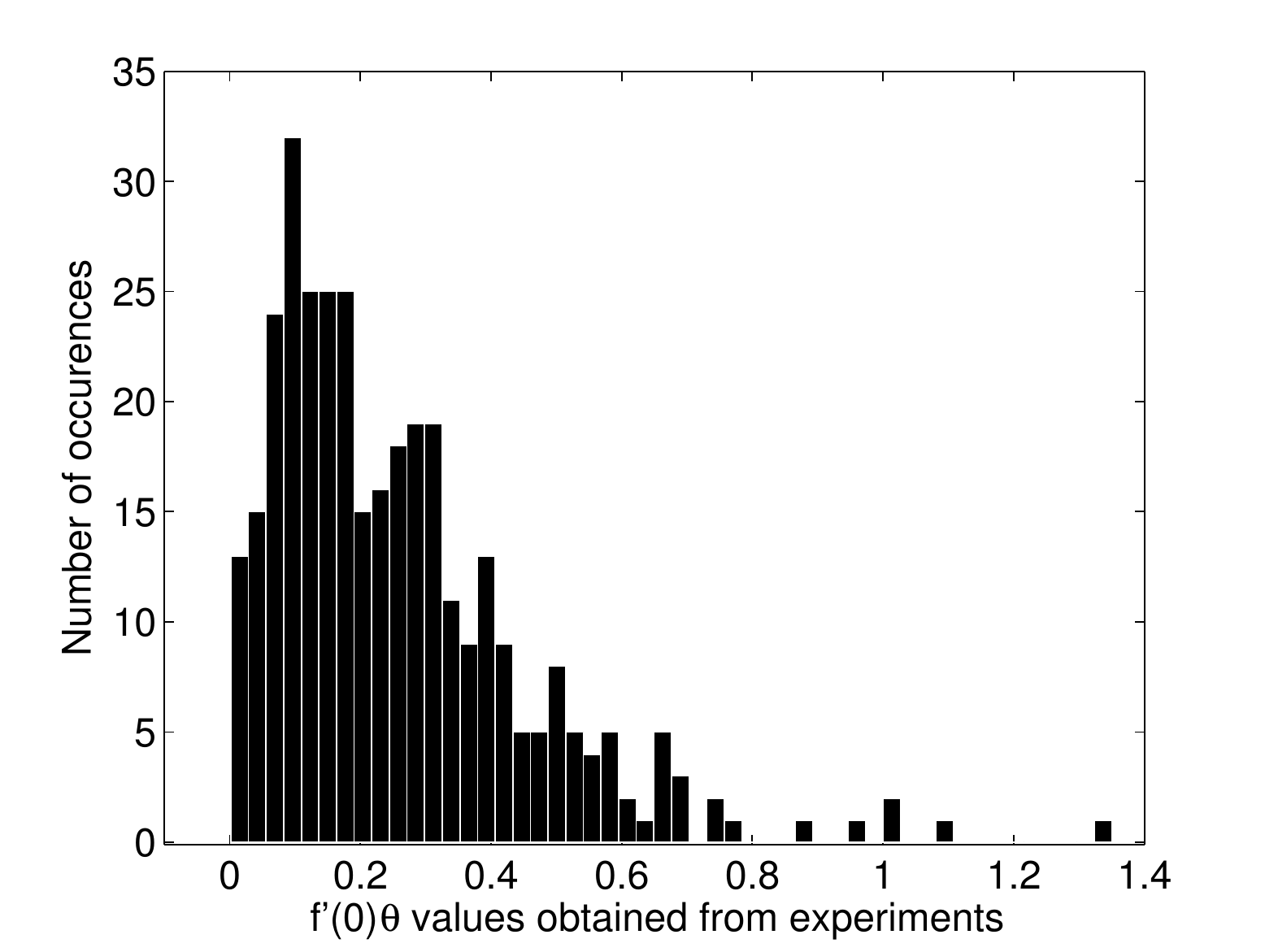}}
\caption{Histogram of the values of the flux factor $f'(0)\theta$ obtained from
experiments. Due to fluctuations in measurements and to the difficulty to detect
very low concentration increments at the beginning of the pulse sequence, 20 out
of 360 experimental $\delta x$ values were negative and thus lead to negative
$f'(0)\theta$. Those values were removed from this histogram.}\label{fig:histo}
%\end{center}
\end{figure}

We checked the approximation $f(\theta)\approx f'(0) \theta$ by fitting straight
lines to the numerically computed $f(\theta)$. As shows Fig.~\ref{fig:flux_2},
the fits are in very good agreement with numerical data. The different plots
correspond to different internal to external conductivitiy ratios
$\sigma_i/\sigma_e$, as indicated in the legend. One can see that the slopes
$f'(0)$ depend on these conductivity ratios, and increase for increasing ratios.
This makes sense, as one would not expect any flux of the electric field for a
perfectly non conductive object. This is also interesting because it gives
another reason why the relative DNA concentration increments seems to increase
during one experiment. This is because the solution inside the GUV being
initially less conductive than the external medium, the electropermeabilization
leads to the mixing of the internal and external solutions, and thus to an
increase of $\sigma_i$ towards $\sigma_e$. This increase of the conductivity
ratio causes in turn an increase of the flux $\phi$, and hence an increase of
$\delta x$. One can also see on Fig.~\ref{fig:flux_2} that for conductivity
ratio values between 1 and 10 corresponding to our experiments, the flux factor
$f'(0)\theta$ should be, for moderate angles, on the order of 0.1--0.5. Its
value should of course be varying from one data point to another, mostly because
of the difference between the permeabilized areas, but our analysis nevertheless
states that if our assumptions are valid, the values of $f'(0)\theta=4\pi
R\delta x / (3 |\mu| E_0 \tau)$ obtained with our experimental data should be
distributed around 0.1--0.5. We have gathered in Fig.~\ref{fig:histo} all our
experimental values of $f'(0)\theta$, using the value of $\mu$ given above. This
histogram contains more than 300 individual data points, corresponding to
various vesicle sizes $R$, field amplitudes $E_0$, and pulse durations $\tau$.
The narrow shape of the distribution around the expected values thus  supports
our hypothesis that DNA entry in the GUVs mostly happens via electrophoretic
effects.

The value of the flux factor  $f'(0)\theta$  averaged on all experiments is
$\langle  f'(0)\theta \rangle\approx0.26$. If we take  the value of
2.34~rad$^{-1}$ for $f'(0)$, corresponding to a conductivity ratio equal to 1
(this has to be true after a few permeabilizing pulses, once the internal and
external solutions have mixed), we obtain for the average value of the
permeabilization angle $\theta$: $\langle \theta \rangle\approx6.3^{\circ}$.
This angle value seems reasonable, but should not be interpreted as the most
frequent permeabilization angle in our experiments. Indeed, this quantity is
extracted from an averaging process on many experiments performed with various
electrical parameters and on GUVs of different sizes. Given the fact that the
vesicles shrank during an experiment, one would expect that $\theta$ decreases
too. Let us recall that in similar systems, single macropores could not be
detected during the pulse application \cite{Por10}. The result of this analysis
is that the experimentally observed 
DNA transfer into vesicles  can be interpreted as, on average, that which would
be given by
electrophoresis through an aqueous pore subtending an average angle of about
$6^\circ$ in the 
direction of the applied field. We see that this conclusion has the merit of
being consistent with 
the assumptions of our model, the effective region across which DNA is
transfered is macroscopic
but does not represent a large fraction of the vesicle surface.   
%%%%%%%%%%%%%%%%%%%%%%%%%%%%%%%%%%%%%%%%%%%%%%%%%%%%%%%%%%%%%%%%%%%%%%%%%%%%%%%%
%%%
%%%%%%%%%%                              CONCLUSIONS                     
%%%%%%%%%%
%%%%%%%%%%%%%%%%%%%%%%%%%%%%%%%%%%%%%%%%%%%%%%%%%%%%%%%%%%%%%%%%%%%%%%%%%%%%%%%%
%%%

\section{Discussion}

% \Delta\psi_c
Before answering the questions \textit{(i)} and \textit{(ii)} introduced in
section~\ref{sec:intro}, about the mechanism of DNA electrotransfer into GUVs
and the ability to control the  amounts transfered, respectively, we shall
briefly discuss our results regarding the decrease in size of the GUVs during
electropulsation. As observed in \cite{por09} the application of electropulses 
leads to a reduction in the size of the vesicles. This reduction in size
eventually stops when the vesicles have a size such that the permeabilization
threshold is no longer exceeded. Our estimation  of the critical membrane
permeabilization potential 
$\Delta\psi_c$ necessary to induce permeabilization turns out to be smaller than
that estimated for the 
same vesicle system (EggPC GUVs) in the absence of DNA. This suggests the
possibility that negatively charged macromolecules may help to induce the
permeabilization process via some interaction with the membrane, which was also
concluded from chemical relaxation spectrometry experiments on small unilamellar
vesicles subjected to an electric pulse in presence of DNA \cite{Fra05}. It is
possible that DNA may flow into the defects that eventually lead to macropores 
and prevent their closing. The presence of DNA in blocked defects could be
related to the endocytosis based electrotransfer mechanism seen in
\cite{che90}. 

% question (i) sur le mécanisme d'entrée
\textit{(i)}
Our qualitative observations and the quantitative agreement of our theory with
our experimental results support  a direct uptake mechanism, as described in
\cite{lur00}, which is  mainly driven by electrophoretic effects. However there
are some differences between our system and protocol and those used by
Chernomordik \latin{et al.} \cite{che90} which should be highlighted. For
example, in \cite{che90} just {one} pulse was applied whereas we used a train of
at least ten pulses. As can be seen in the top left panel of
Fig.~\ref{fig:fluo}, the amount of DNA transferred after a single pulse could
sometimes be very small. We could infer that in fact no DNA penetrated the
vesicle during the pulse and that the endocytosis-like mechanism, as we observed
in some rare cases, occured after the electrical treatment. The intensities of
the field  applied in \cite{che90} induced transmembrane voltages of the order
of at most 450 mV  whereas ours, and those used by Lurquin and Athanasiou,
caused initial potential drops of more than 1 V. This could be an explanation
for  why so little DNA  has been taken up by the liposomes in the experiments of
\cite{che90}. Another origin for the difference of the reported behaviours could
lie in the nature of the systems themselves.  The liposomes used in \cite{che90}
were much smaller (LUVs of $\approx 500$ nm diameter) and contained 70 \% DPPC
and 30 \% cholesterol, whereas the DPPC GUVs of \cite{lur00}, and ours made of
EggPC, did not contain cholesterol.
DNA interaction with the membrane of CHO cells during electrotransfection
experiments appears to be more complex than that inferred here. In \cite{gol02}
the initial interaction between DNA and the CHO cell membrane leads to the
formation of spots on the cell surface, again facing the cathode, where DNA
aggregates are formed. The fact that this spot formation is not seen in GUVs has
many possible explanations. For DNA and cells it is possible that there is a
physical or chemical interaction which tends to trap DNA near the surface or it
could be that the reduced mobility of DNA in  the cell interior causes a traffic
jam like phenomenon where the DNA is locally blocked \cite{Esc10}.   It is worth
mentioning that the asymmetric transport phenomena discussed below  also occurs
in real cells \cite{tek94}. The asymmetric transport pattern found by the
authors led them to suggest that pores were created on both sides of the
membrane, but with a larger size (and also lower number) on the cathode-facing
hemisphere.

% question (ii) sur la possibilité de contrôler les quantités transférées
\textit{(ii)}
The agreement of our experimental results with our model for DNA uptake
corroborates our quantitative observations. It means that we can indeed control
and predict the quantity  of electrotransfered DNA per pulse $\delta x$. Even if
we can not accurately monitor the size, shape, number and nature of the membrane
defects allowing DNA entrance, we have seen that assuming an entry through a
macropore of angle $\sim 6^\circ$ (or equivalently a flux factor $f'(0)\theta$
in Eq.~\ref{eqn:deltax} on the order of 0.2) provides a satisfactory prediction
of the transfered amounts. Although no macropore could be detected \emph{during}
the pulse application \cite{Por10}, this simplification turns out to be a useful
and convenient assumption for estimating $\delta x$.
Furthermore, as was observed in living cells \latin{in vitro} \cite{Kle91,Suk92}
and in skeletal muscle \latin{in vivo} \cite{Bur00}, DNA uptake by the GUVs
indeed appears to be dominated  by the integrated effect of the electrophoretic
force resulting from the electric field application, {\em i.e.}  the term $|\mu|
E_0 \tau$ in Eq.~\ref{eqn:deltax}. The effect of the electric pulses is thus
twofold: they both permeabilize the lipid bilayer, and push the charged
compounds inside the liposome, making the electrotransfer of highly negatively
charged macromolecules very efficient. This importance of the electrophoretic
effect is reminiscent of the observation that electric fields generated by salt
gradients across artificial nanopores enhance the capture rate of DNA molecules
into the pores \cite{Wan10}.
Another significant experimental finding is the following. As can be seen on the
bottom right panel of Fig.~\ref{fig:fluo}, we could sometimes  reach DNA
concentrations inside the vesicle greater than the concentration outside
($c/c_0=I/I_0>1$). The inverse dependence of $\delta x$ on the vesicle radius
$R$ alone cannot  be  responsible for a concentration build up within vesicles
exceeding that of the bulk. Indeed, even if $R$ decreases during an experiment
(and thus $\delta x$ increases, see Eq.~\ref{eqn:deltax}), one should not in
principle be able to exceed the outer concentration $c_0$. The build up of an
excess concentration strongly suggests that DNA can not cross the membrane on
the anode facing pole, and means that the permeant structures reported to be
created on this side \cite{tek94,tek01} have sizes on the order of a few nm at
most. It also means that the pore created on the other side closed relatively
quickly at the end of the pulse, otherwise the inner concentration would always
have time to equilibrate with the outer one  (as the osmotic gradient will
dominate in the absence of an applied field). Indeed, it had been reported that
electropores induced by DC pulses in GUVs resealed within some tens
\cite{ris05}, at most a few hundreds of ms \cite{Por10} for similar electric
pulses, fast enough to prevent significant DNA concentration changes due to
diffusion. The presence on the anode facing pole of the membrane tubules
reported previously in \cite{por09} and which are also observed with EggPC
vesicles (see Fig.~\ref{fig:photo}) could presumably also prevent the DNA from
leaving the vesicle from that pole.
 An interesting corollary  of these  results  would be that electromediated
loading of 
vesicles with positively charged macromolecules should be much less efficient
than that observed here.
This is because positively charged macromolecules would be forced away from the
side of the 
vesicles containing the macropores.

% Recette et vente de la méthode.
Besides  clarifying   the mechanism of DNA uptake by vesicles, our work
demonstrates the efficiency of the loading  method. These results should be
relevant to the encapsulation of  plasmid DNA into giant liposomes, for the
purpose of gene transfection for example. A simple rule of thumb for choosing
the electric field parameters for a loading protocol would be: use 
a pulse duration of $\tau=5$ ms, tune the field amplitude $E_0$ according to the
vesicle initial size $R_0$ in order to obtain induced transmembrane voltages
$\Delta\psi_0=(3/2)R_0E_0\sim 1$~V, and then apply a sequence of pulses at 0.33
Hz repetition frequency until the required DNA concentration inside the vesicle
is attained.  This loading technique is also applicable to other negatively
charged macromolecules, and an equivalent efficiency ($c/c_0>1$) could be
attained provided the molecules are big enough to be unable to cross the
membrane at the anode-facing side. The higher the electrophoretic mobility, the
higher the electrotransfered amount, as can be understood from Eq.~\ref{eqn:deltax}; thus this technique would be particularly efficient with
highly charged molecules.
The negative charge is crucial in order to reach the vesicle from the
cathode-facing side; indeed, positively charged molecules would reach the GUV
from the anode-facing side, which hosts the nucleation of the tubular structures
shown in Fig.~\ref{fig:photo}, and whose crossing is much more difficult, as
shown by the fact that we oberved surconcentrations of DNA ($c/c_0>1$).

\section{Conclusion}
In this paper we have described a quantitative method for loading GUVs with
negatively charged macromolecules, and we have shown that the predominant
pathway of electromediated DNA uptake into liposomes is undoubtedly the
electrophoretic entrance in a  free form \latin{via} defects created on the
cathode-facing pole of the vesicles. This spectacular symmetry breaking, first
observed on GUVs in \cite{tek01}, is what makes electromediated DNA uptake by
vesicles so efficient. Indeed, in some of our experiments, we can reach DNA
concentrations inside the vesicles higher than external concentrations. This
would not have been possible if large pores also opened on the anode-facing
hemisphere (as  DNA would flow out through these pores), and clearly shows that
even if the vesicle is permeabilized to small compounds at the anode side,
sufficiently large moelcules cannot cross the membrane in this region. From this
we can infer that the permeant structures created opposite  the positive
electrode have sizes of the order of a few nanometers at most. The underlying 
origin of these striking differences of the membrane reorganization at the
moleculer level are not yet understood, and explaining this asymmetry represents
an exciting direction for further research.

%%%%%%%%%%%%%%%%%%%%%%%%%%%%%%%%%%%%%%%%%%%%%%%%%%%%%%%%%%%%%%%%%%%%%%%%%%%%%%%%
%%%
%%%%%%%%%%                      ACKNOWLEDGEMENTS                        
%%%%%%%%%%
%%%%%%%%%%%%%%%%%%%%%%%%%%%%%%%%%%%%%%%%%%%%%%%%%%%%%%%%%%%%%%%%%%%%%%%%%%%%%%%%
%%%

\section*{Acknowledgements}
This work was partially supported by the Institut Universitaire de 
France. We wish to thank Luc Wasungu for his help with DNA purification and also
\'Emilie Phez for her 
participation in the inception of this work. TP, CF, JT and MPR are members of
the CNRS consortium CellTiss; CF is also a member of the CNRS consortium
Microscopie Fonctionnelle du Vivant. Imaging experiments were run at the TRI
platform.

\balance

\footnotesize{
\bibliography{biblio_guv_dna} %your .bib file
\bibliographystyle{rsc} %the RSC's .bst file
}

\end{document}